# Non-volatile Phase-only Transmissive Spatial Light Modulators

Zhuoran Fang[1,+,*], Rui Chen[1,+,*], Johannes E. Fröch[1,2], Quentin A. A. Tanguy[1], Asir Intisar Khan[3], Xiangjin Wu[3], Virat Tara[1], Arnab Manna[2], David Sharp[2], Christopher Munley[2], Forrest Miller[1,4], Yang Zhao[1], Sarah J. Geiger[4], Karl F. Böhringer[1,5], Matthew Reynolds[1], Eric Pop[3], Arka Majumdar[1,2*]

[1]*Department of Electrical and Computer Engineering, University of Washington, Seattle, WA 98195, USA*

[2]*Department of Physics, University of Washington, Seattle, WA 98195, USA*

[3]*Department of Electrical Engineering, Stanford University, Stanford, CA 94305, USA*

[4]*The Charles Stark Draper Laboratory, Cambridge, MA 02139, USA*

[5]*University of Washington, Institute for Nano-engineered Systems, Seattle, Washington, United States*

[+]*These authors contributed equally*

*Email: arka@uw.edu, rogefzr@uw.edu, charey@uw.edu

## Abstract

Free-space modulation of light is crucial for many applications, from light detection and ranging to virtual or augmented reality. Traditional means of modulating free-space light involves spatial light modulators based on liquid crystals and microelectromechanical systems, which are bulky, have large pixel areas ($\sim 10 \mu m \times 10 \mu m$), and require high driving voltage. Recent progress in meta-optics has shown promise to circumvent some of the limitations. By integrating active materials with sub-wavelength pixels in a meta-optic, the power consumption can be dramatically reduced while achieving a faster speed. However, these reconfiguration methods are volatile and hence require constant application of control signals, leading to phase jitter and crosstalk. Additionally, to control a large number of pixels, it is essential to implement a memory within each pixel to have a tractable number of control signals. Here, we develop a device with nonvolatile, electrically programmable, phase-only modulation of free-space infrared radiation in transmission using the low-loss phase-change material (PCM) $Sb_2Se_3$. By coupling an ultra-thin PCM layer to a high quality (Q)-factor ($Q\sim406$) diatomic metasurface, we demonstrate a phase-only modulation of $\sim0.25\pi$ ($\sim0.2\pi$) in simulation (experiment), ten times larger than a bare PCM layer of the same thickness. The device shows excellent endurance over 1,000 switching cycles. We then advance the device geometry, to enable independent control of 17 meta-molecules, achieving ten deterministic resonance levels with a 2π phase shift. By independently controlling the phase delay of pixels, we further show tunable far-field beam shaping. Our work paves the way to realizing non-volatile transmissive phase-only spatial light modulators.

**Introduction**

Free-space modulation of light is a key enabling technology behind optical communications, holography, ranging, and virtual/ augmented reality. Traditional spatial light modulators (SLMs) based on liquid crystal (LC) on silicon[1] or micro-electro-mechanical systems (MEMS)[2] employ large pixels ($\sim 10\mu m \times 10\mu m$), resulting in bulky devices and generally require large driving voltage. To address these limitations, recent years have seen tremendous effort to realize free-space light control based on subwavelength diffractive optical metasurfaces integrated with active materials[3–12]. Metasurfaces can support resonances that enable substantial phase or amplitude modulation with small pixel-size. Since a smaller active volume is modulated, lower energy and faster modulation speed can be achieved. For example, metasurface resonators ($Q\sim 550$) based on organic electro-optic (EO) polymers have enabled GHz modulation speed[13]; LCs combined with a Huygen's metasurface significantly reduced the pixel size (~1μm) and LC thickness (~1.5μm) required to attain a full 2π phase shift range[3]; metasurfaces based on plasmonic resonances coupled to epsilon-near-zero materials have enabled a full 2π modulation with independent control of amplitude and phase[4]; a large phase-modulation of ~1.3π was reported by tuning a plasmonic metasurface using graphene[12]. Nevertheless, these approaches are all based on volatile changes such as the Pockels effect or free carrier dispersion, necessitating a constant power supply to hold the static state. To ensure time-multiplexed pixel control, most of these volatile SLMs require an active memory matrix with transistors to hold the written state. In addition, the most prevalent LC-based phase-only SLMs suffer from detrimental pixel crosstalk[14] and temporal phase fluctuations or jitter[15]. To overcome some of these limitations, a promising solution is to modulate light using non-volatile materials, which can drastically improve energy efficiency and avoid phase flickering. Moreover, the control complexity can be significantly reduced due to the built-in memory of the non-volatile reconfiguration. Chalcogenide-based phase-change materials (PCMs) are ideal candidates to realize such functionality[16], thanks to their non-volatile microstructural phase transition[17], large contrast in complex refractive index (typically $\Delta n \geq 1$)[18], and CMOS compatibility[19]. Since PCMs hold their state once configured, a truly 'set-and-forget' switching element can be realized. In fact, PCMs have already attracted considerable attention to create tunable metasurfaces for applications such as varifocal lensing[8,20], beam steering[21,22], intensity switching[6,7,23,24], and spectral filtering[25]. Despite the progress, non-volatile optical phase-only modulation in transmission - a highly desirable feature for SLMs - remains elusive. This is because previous works on nonvolatile tunable metasurfaces have used lossy PCMs such as GST[6,24] or GSST[7] which concomitantly induce large absorption upon phase transition, prohibiting optical phase-only modulation. The use of metallic heaters further leads to ohmic losses and makes transmissive operation difficult. In fact, to date, all electrically tunable PCM meta-optics have been demonstrated in reflection mode. Although non-volatile phase-only control has been demonstrated in mid infrared where GST is transparent[26], the PCM was switched optically, which prevents a fully integrated and scalable platform. A similar work has shown near $2\pi$ phase shift via optical switching of low-loss PCM $Sb_2S_3$ in the visible spectrum[27], but phase modulation is accompanied by a large amplitude change due to non-negligible loss of crystalline $Sb_2S_3$.

By leveraging a thin layer of low-loss PCM $Sb_2Se_3$[28], we demonstrate a non-volatile, electrically programmable metasurface for phase-only modulation of free-space near infrared (NIR) light in transmission. The ultra-low loss of $Sb_2Se_3$ enables decoupling of the phase modulation from amplitude modulation in NIR. A phase-modulation of ~$0.25\pi$ in simulation and ~$0.2\pi$ in experiment is achieved by coupling the PCM to a high-$Q$ diatomic metasurface ($Q \sim 409$). The tunable metasurface also demonstrates large endurance with over ~1,000 transitions. We further exploited a guided-mode resonance, albeit with a lower $Q$-factor, to facilitate a larger electric field overlap with the $Sb_2Se_3$ layer, which enables a full $2\pi$ phase shift. Beyond global control of the entire metasurface, we further show independent electrical control of 17 meta-molecules, verified by optical microscope images and the spectral response, which shows discrete and reversible resonance levels as function of the respective configuration. A deterministic multi-level resonance tuning of the metasurface is achieved by switching the meta-molecules one-by-one with a total resonance shift of ~8nm at a center wavelength of ~1230nm.

By imparting different phase profiles through independently controlling individual pixels, we demonstrate dynamic beam focusing with three distinct focal distances. This work constitutes a crucial step towards a truly non-volatile "set-and-forget" transmissive phase-only SLM.

**High-$Q$ silicon diatomic metasurfaces**

$Sb_2Se_3$ undergoes a refractive index contrast of ~0.7 and exhibits zero loss in both states[28,29] near 1550 nm upon phase transition, which stipulates only a ~2.2 µm PCM thickness to obtain a $2\pi$ phase difference. Although this is significantly thinner than for LCs used in commercial SLMs (> 3 µm), it still poses a significant challenge in reversible switching PCMs. As melt-quenching is required to amorphize the material, the integration of thick materials (~ 50 nm[30]) is precluded, as this limits the required critical cooling rate[31]. Motivated by the use of microring resonators to increase the modulation contrast in integrated photonics[32], a high-$Q$ meta-optical resonator can be used to enhance the phase-modulation of free-space light by thin film PCMs, while allowing thinner device layers than traditional Fabry-Perot cavities. Earlier works on high-$Q$ meta-optical resonators focused on high contrast gratings (HCGs)[33], whereas recent works studied quasi-bound-state-in-the-continuum (q-BIC) in periodic nanostructures with in-plane asymmetry[34,35]. Here we combine the idea of HCGs and q-BIC by introducing an asymmetry into the periodicity of traditional HCGs, realizing a diatomic metasurface (Fig. 1a). Although such diatomic metasurfaces have been explored in theory[36–38], thus far the experimental demonstration of this concept has been missing. Fig. 1a shows how the symmetry of the HCG is broken by slightly displacing one of the gratings relative to the other such that the grating spacing becomes dissimilar i.e., $a_1 \neq a_2$. We denote the displacement or perturbation by $\delta = \frac{a_2-a_1}{\Lambda}$ where $\Lambda$ is the period of the diatomic grating, with $\delta = 0$ denoting a simple HCG. The duty cycle $\Gamma$ is defined as $\frac{w}{\Lambda_{HCG}}$ for HCGs and $\frac{2w}{\Lambda}$ for diatomic gratings, where $w$ is the grating width. We note that, $\Gamma$ is essentially identical for both types of gratings because the period of diatomic gratings is twice that of the HCG. The periodicity doubling causes the folding of the first Brillouin zone such that dark modes that were originally at the edge of the Brillouin zone outside the light cones, get folded into the interior of the light cones, allowing for free-space excitation[36,37]. Fig. 1b shows the simulated spectral response of HCGs (δ=0) and diatomic gratings (δ=0.05) upon excitation of normal incident transverse-magnetic (TM) polarized light near 1550nm wavelength (Λ=900nm and Γ=0.7). When δ=0, the dark modes are not accessible to the free-space excitation and hence no resonance is observed. While a small perturbation of δ=0.05 is enough to introduce a sharp q-BIC resonance[36] with its signature Fano line shape. The pronounced resonance is also clearly visible in Supplementary Fig. S1, where the q-BIC resonance represents ∼40 × electric field enhancement compared to the δ=0 case. By fitting the experimentally measured resonance, we extract a large $Q$ factor of 409 for heavily N-doped 220nm-thick silicon-on-insulator (SOI), as shown by the inset of Fig. 1b. We note that the doping is necessary for the electrical actuation of the phase transition in the PCM[39,40]. Such a high experimental $Q$ factor is attributed to the q-BIC mode and the ultra-smooth etching side wall as revealed by Fig. 1c, despite the additional loss from the doping. Fig. 1d shows the experimentally measured resonances as the period increases from 900nm to 1000nm (silicon thickness of 220nm, δ=0.05 and Γ=0.7). The resonance almost linearly shifts across the telecommunication wavelength range without significant change in the $Q$ factor – a highly desirable characteristic of diatomic gratings[36]. In contrast, changing the period of HCGs will lead to a dramatic change in the $Q$ factor as reported previously[33].

The diatomic metasurface makes an ideal platform for demonstrating non-volatile phase-only modulation with low-loss PCMs for two reasons. First, meta-atoms of equal width $w$ allow identical microheater resistance, which is essential for uniform heating of PCMs cladded on top. In comparison, meta-atoms of dissimilar width or with additional notches[41] can also facilitate high $Q$ resonances, but will ultimately give rise to different resistances in the heater resulting in a nonuniform Joule heating from current injection. Secondly, the resonance can be fine-tuned to an arbitrary wavelength by changing the period without degrading the $Q$ factor, similar to changing the radius of a microring in integrated photonics, which enables operation over a large wavelength range limited only by the material bandgap.

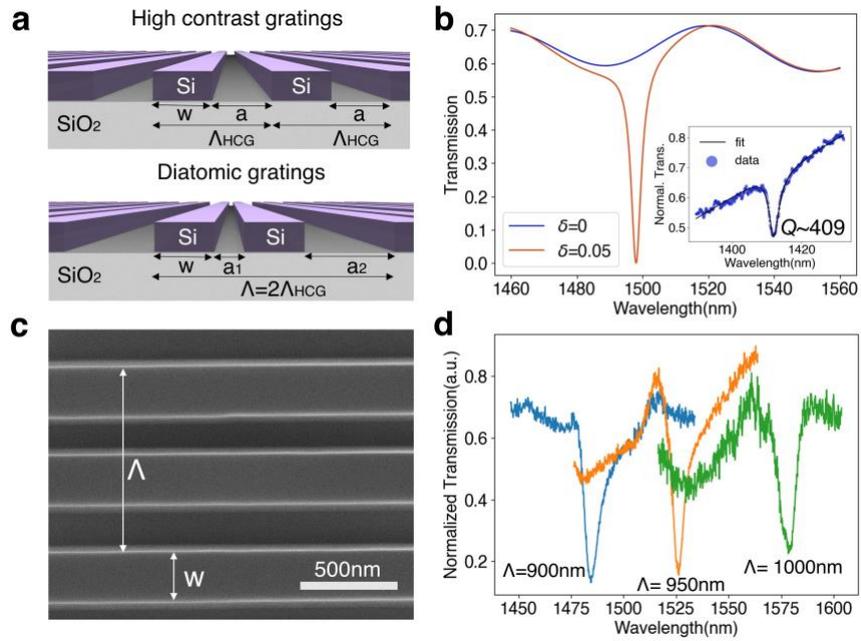

**Figure 1: High-$Q$ silicon diatomic metasurfaces**

**a** Schematics of the high contrast gratings (top) and diatomic gratings (bottom). $\Lambda_{HCG}$ and $\Lambda$ are the periods of the high contrast gratings and diatomic gratings respectively. The spacing between the gratings are denoted as $a_1$, and $a_2$. $w$ is the grating width. **b** Simulated transmission spectrum of the high contrast gratings (blue) and diatomic gratings (orange) under normal incident TM polarized light. Silicon thickness is 220nm, $\Lambda$=900nm and $\Gamma$=0.7. Inset: Experimental spectrum of a resonance fitted by Fano equation, $\Lambda$=950nm and $\Gamma$=0.7. **c** Scanning electron micrograph (SEM) of fabricated silicon diatomic gratings. **d** Measured spectra of diatomic gratings with different periods from 900nm to 1000nm, with $\delta$=0.05 and $\Gamma$=0.7.

**Non-volatile electrically programmable metasurface for transmissive phase-only control**

In order to dynamically control the diatomic metasurface, we dope the SOI to create microheaters [39,42] (phosphorus doping concentration ~$10^{20}$ cm$^{-3}$), that can switch the Sb$_2$Se$_3$ cladded on top (Fig. 2a). This doped silicon is then etched into diatomic metasurface, before depositing 20nm Sb$_2$Se$_3$ via sputtering, followed by 40nm atomic-layer-deposited (ALD) Al$_2$O$_3$ encapsulation to prevent oxidation and PCM dewetting during switching (See Methods for details on fabrication). Ohmic contacts are then formed by Ti/Au electrodes. The cross-section of the metasurface is shown in Fig. 2a(ii). Current is injected into the highly doped silicon gratings via electrical pulses that causes joule heating, which in turn switches the PCM. A two-objective transmission setup (see Methods for details on the measurement setup) is used to probe the reversible switching of Sb$_2$Se$_3$-loaded diatomic metasurface. The fabricated chip is wire bonded to a customized printed circuit board (PCB) (labelled PCB1 in Fig. 2b) connected to a second customized PCB carrying a microcontroller (labelled PCB2 in Fig. 2b). PCB2 is then connected to an arbitrary function generator (AFG). The microcontroller on the PCB2 can be programmed to individually address each metasurface pixel, or meta-atoms (see Methods for details on the electrical control). Fig. 2c shows the optical micrograph of a 30μm aperture metasurface on a chip that has been wire bonded to the PCB. Reversible tuning of the diatomic resonance is shown in Fig. 2d (simulation) and Fig. 2e (experiment). The spectra of three consecutive switching cycles are plotted in Fig. 2e and the shaded regions indicate standard deviations between the cycles. The small standard deviation reveals excellent cycle to cycle reproducibility. The experimentally extracted spectral shift ($\Delta\lambda \sim 1.2 nm$) matches very well with the simulated shift ($\Delta\lambda \sim 1.3 nm$). Fig. 2f shows the phase shift ($\Delta\phi$) and transmission contrast ($\Delta T\%$) between the two optical states near the 1518nm resonance. A good agreement between the simulation and the experiment can be clearly observed. We extract a maximum phase shift of $\sim 0.2\pi$ ($\sim 0.25\pi$) near the resonance wavelength via digital holography experiment (simulation) with less than 10% measured change in transmission (see Methods for details on the spectra and phase-shift measurement). We note that, this phase shift is ~10 times larger than what can be achieved by switching a 20nm thick blanket film of $Sb_2Se_3$ without a metasurface. To further show that this is indeed a phase-only modulation, we fit the Fano line shape to the resonances of $aSb_2Se_3$ and $cSb_2Se_3$ averaged over five consecutive cycles (Supplementary Fig. S2). The extracted $Q$ factor decreases from 312 to 271 upon crystallization, which shows that minimal loss is introduced by the phase transition. In contrast, the resonance is completely suppressed by the same thickness of GST cladded on the metasurface due to high absorption in GST (Supplementary Fig. S3). Temporal trace measurements (Supplementary Fig. S4) also confirm that the change is indeed non-volatile and hence caused by the $Sb_2Se_3$. Finally, we show that the tunable metasurface maintains its functionality and is robust for over 1,000 switching events, without degradation in contrast, as shown in Fig. 2g, where a tunable laser is set to a wavelength near the resonance to detect a measurable change of ~ 6% ± 1% in transmission (See Methods for details on the cyclability test). Furthermore, optical micrographs taken before and after the cyclability test indicate no ablation occurred by the switching across the entire

$30 \mu m \times 30 \mu m$ metasurface (Supplementary Fig. S5). Apart from switching Sb$_2$Se$_3$, we also show that the doped silicon microheater is a versatile platform for switching other PCMs, for example we have also reversibly switched GST using a doped silicon microheaters on sapphire substrate to realize a broadband tunable notch filter in transmission, see Supplementary Fig. S6.

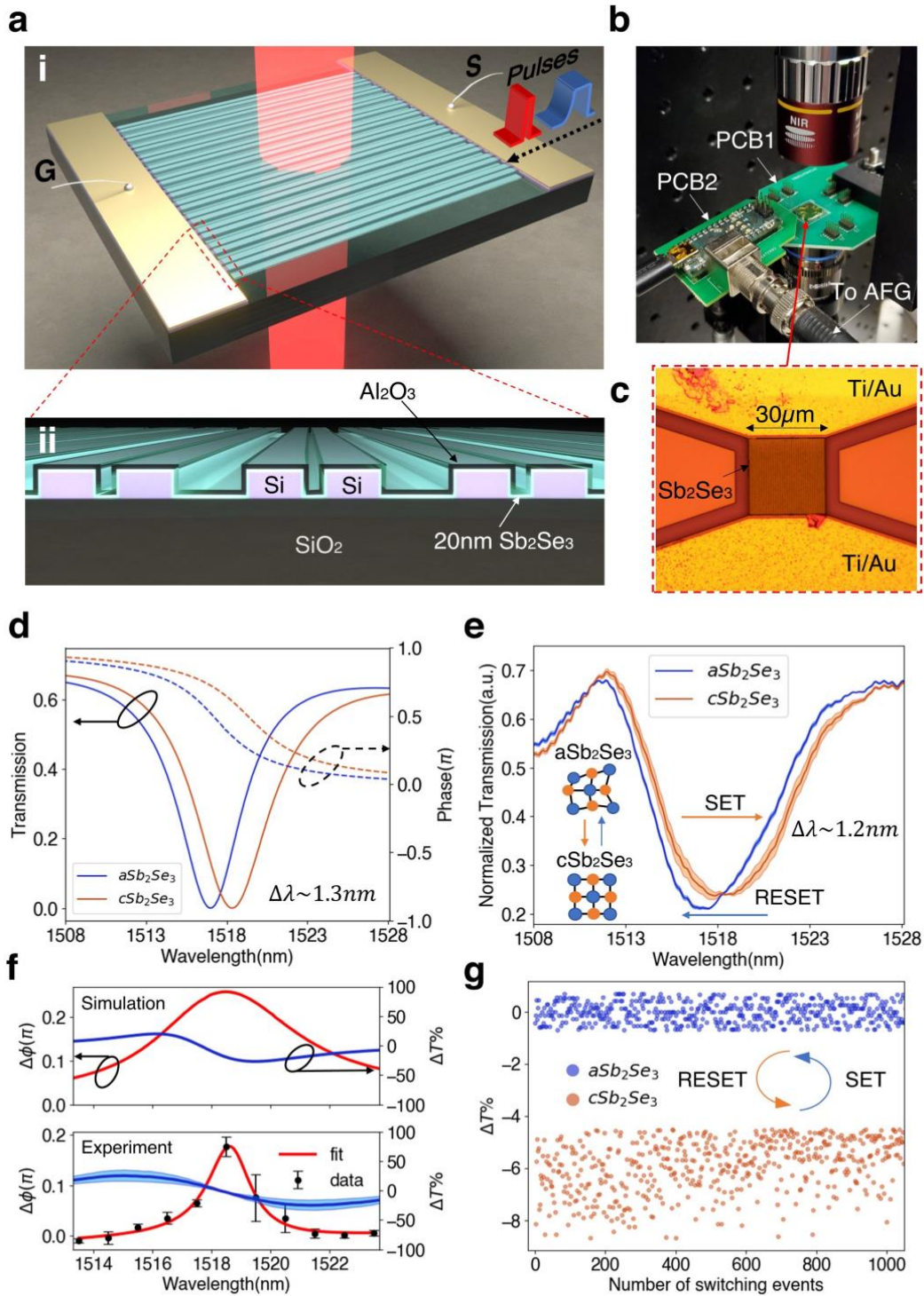

**Figure 2: A nonvolatile electrically reconfigurable metasurface based on $Sb_2Se_3$**

**a** Schematic of the transmissive tunable diatomic metasurface based on $Sb_2Se_3$. **(i)** Perspective view. S (G), signal (ground) electrode. **(ii)** Zoomed-in cross-sectional view. **b** Optical setup for probing the metasurface in transmission along with the devices under test wire bonded to a customized PCB. AFG is arbitrary function generator. **c** Optical micrograph of a single metasurface on the chip under test. **d** Simulated spectral and phase shift caused by the phase transition of 20nm thick $Sb_2Se_3$. $\Delta\lambda$ is the wavelength shift of the resonance dip. The diatomic metasurface is designed to have $\Lambda$=900nm, $\Gamma$=0.7, and $\delta$=0.05. a(c)$Sb_2Se_3$: amorphous(crystalline) $Sb_2Se_3$. **e** Measured reversible switching of the diatomic resonance. The switching conditions are 3.6V, 50μs pulse width, 30μs trailing edge for SET and 11.6V, 1μs pulse width, 8ns trailing edge for RESET. Three consecutive cycles are plotted where the shaded area indicates the standard deviation between the cycles and the solid line indicate the average. The spectrum is normalized with that of bare $SiO_2$ on silicon. **f** Phase shift ($\Delta\phi$, red) and transmission contrast ($\Delta T\%$, blue) between two optical states extracted from simulation (top) and experiment (bottom). The phase is measured at 11 different wavelengths with 1nm spacing and averaged over three switching cycles. The standard deviation over the cycles is shown by the error bars. **g** Cyclability of the tunable metasurface for 1,000 switching events. The switching conditions are 3.6V, 50μs pulse width, 30μs trailing edge for SET and 11.6V, 600ns pulse width, 8ns trailing edge for RESET. Each pulse is temporally separated by 2s to ensure long thermal relaxation. The data is filtered by a two-point moving average to reduce fluctuation caused by thermal and mechanical noises.

**Full 2π phase shift and individual control of meta-molecules**

We further exploit a transverse-electric (TE) polarized guided-mode resonance with enhanced electric field interaction with $Sb_2Se_3$ to achieve a larger resonance shift (~10 nm) and a full 2π phase shift in simulation (Supplementary Fig. S7). We note that this resonance mode requires a slight incident angle (~3 degrees) to couple light to the metasurface (Supplementary Fig. S7). This different resonance provides a much larger resonance and optical phase tuning range, albeit at the expense of a lower *Q*-factor (~100) than the q-BIC designs.

By now we have established global phase-only modulation in transmission of the entire metasurface, whereas spatial light modulation requires independent control of each pixel. Here we demonstrate the independent addressing of meta-molecules (or elements), defined as two diatoms with a total pitch of 1.8μm, to control the optical wavefront spatially and spectrally. The individual control is realized using a single source channel and 17 separate ground channels connected to 17 meta-molecules, see Figs. 3a and 3b. The channels fanned out from the metasurface are electrically isolated by over-etching the silicon into the buried oxide and are wire-bonded to PCB 1, as shown in Fig. 2b. We note that the number of channels used here is arbitrary and is only limited by our PCB design. The source channel delivers the voltage pulses to all ON channels with an equal resistance (Supplementary Fig. S8) to allow simultaneous control. Each ground channel is switched ON/OFF by a field-effect transistor switch on PCB 2, which controls the voltage drop on each meta-molecule.

We implement arbitrary phase masks via the following two-step sequence: (1) first, all meta-molecules are electrically switched to the crystalline phase; (2) a single amorphization pulse is then applied to switch the selected meta-molecules. Amorphization, instead of crystallization, is used for the independent control to avoid channel crosstalk. We observed large crosstalk if crystallization is used to switch the individual channel (Supplementary Fig. S9), as the longer crystallization pulses lead to a more severe thermal dispersion (Supplementary Fig. S10). Additionally, the lower temperature threshold for crystallization compared to amorphization (200 °C for crystallization vs. 620 °C for amorphization) can also contribute to the crosstalk. Using this two-step switching method, four different configurations of the metasurface were obtained, shown in the IR camera images in Fig. 3c: (i) all amorphous metasurface; (ii), (iii) hybrid metasurface with an amorphous period two and three times the original channel pitch respectively, and (iv) all crystalline metasurface. Supplementary Video 1 shows the reversible switching between these configurations. The insets at the bottom left corner of each subplot show a schematic for the respective metasurface configurations for clarity. The bright (dark) lines correspond to amorphous- (crystalline-) $Sb_2Se_3$ from the slight absorption difference. The patterns are implemented by the same phase transition condition (15 V, 1.25 μs for amorphization and 6 V, 50 μs for crystallization), showcasing a universal method to set arbitrary patterns. Besides the IR camera images, we also measured the spectra of the configurations in Supplementary Fig. S11, which qualitatively matches with simulations. Apart from collectively switching multiple channels, we further showed that the channels can be independently addressed one by one (Supplementary Video 2).

Increasing the number of amorphized channels causes a gradual spectral shift of the resonance from 1235 nm to 1228 nm, as shown in Fig. 3d. We fitted the resonance to a Fano line shape and extracted the resonance shift versus the number of meta-molecules switched in Fig. 3e. The small error bars averaged over nine repeated cycles suggest a highly deterministic multilevel operation. Compared to the partial amorphization of a single pixel via pulse amplitude or width modulation, individual tuning of the meta-molecules is much more reliable since each meta-molecule undergoes a full amorphization process. The resonance shift is not observed when more than ten meta-molecules were switched because the focused probe beam with a natural Gaussian shape has low intensity at the edge. The fitted *Q* factor (Supplementary Fig. S12) first decreases from ~120 to ~90 until 4 meta-molecules are switched, and then increases to ~120. We attribute the lower *Q*-factor to inhomogeneous broadening, resulted from the coexistence of the amorphous and crystalline resonance mode. As more meta-molecules are switched, the amorphous resonance gradually dominates, leading to an increase of the *Q*-factor.

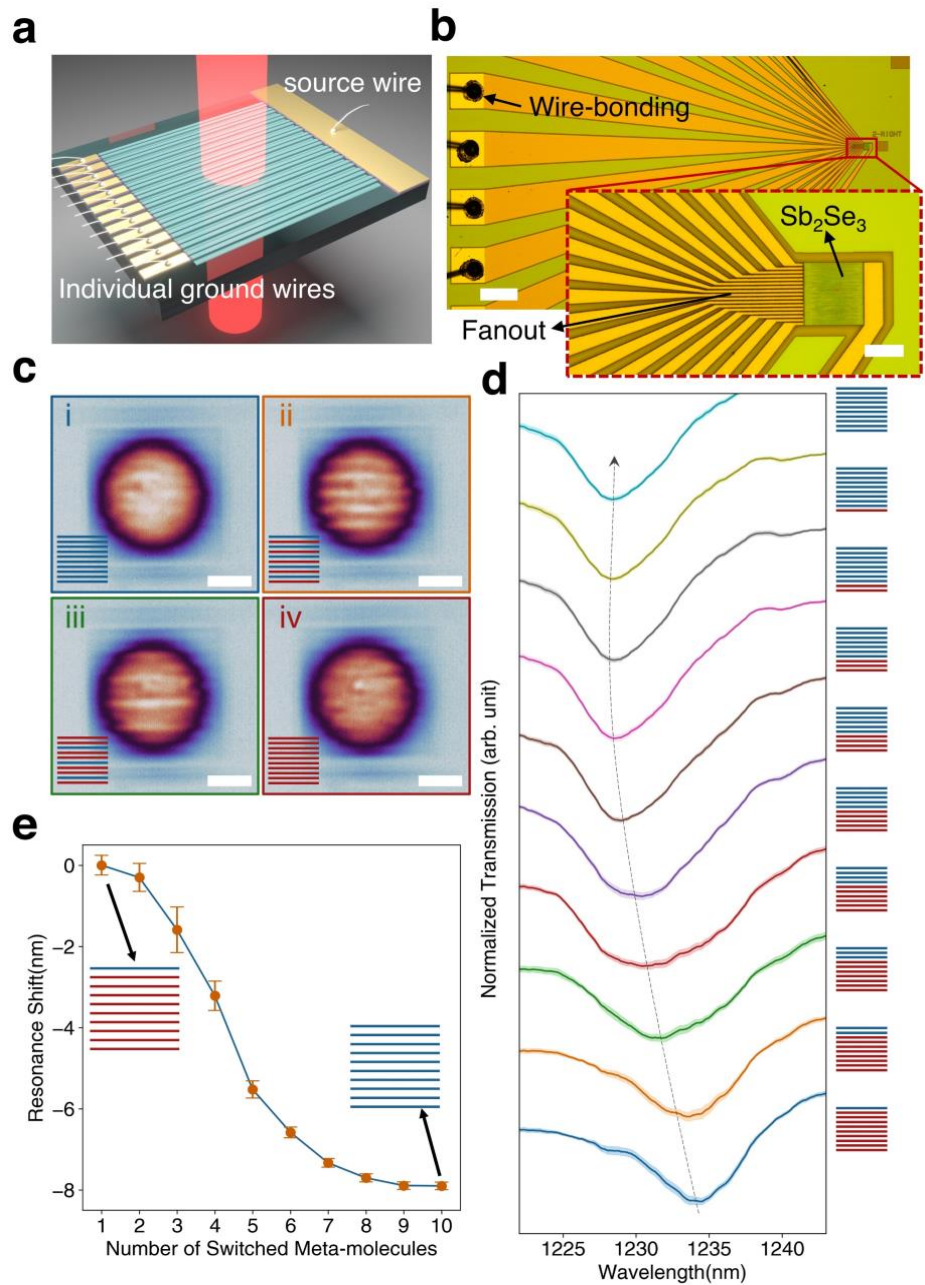

**Figure 3: Individual control of meta-molecules.**

**a** Schematic of the individually controlled meta-molecules. **b** Micrographs of the metasurface with individual meta-molecule controllability (scale bar: 200μm), where each channel is wire-bonded to contact on PCB1. Inset: Zoomed-in micrograph (scale bar: 20μm) shows the fanout of 17 electrical channels from a single metasurface. **c** IR camera images showing four different metasurface configurations (scale bar: 10μm), indicated by the insets at the bottom left corner of each subplot. The circular shape is due to the finite area illuminated by the focused Gaussian beam. The bright (dark) lines show an amorphous (crystalline) $Sb_2Se_3$ due to the slight absorption difference. **d** Waterfall plot for the normalized transmission spectra as the meta-molecules are switched individually. The result is averaged over nine repeated cycles, and the shaded region indicates the standard deviation. Each spectrum is vertically offset for clarity. **e** Fitted Fano-resonance spectral shift with respect to the number of channels switched. The resonance wavelength gradually blue shifts as more gratings are switched. The small error bar indicates a deterministic multilevel operation. (arb. unit: arbitrary unit)

Finally, we demonstrate the device's functionality as a fully configurable spatial light modulator by independently controlling the individual meta-molecules to locally impart phase shifts and to manipulate the transmitted light field. This enables tunable beam focusing at different depths as different numbers of meta-molecules are switched to the amorphized state. Thanks to the significant resonance shift (~10nm), the switching induces a local phase shift near the resonance wavelength (0 ~ $2\pi$, dependent on the wavelength). The simulation result in Fig. 4a shows a distinct change of the intensity profiles along the optical axis and achieve focusing of the transmitted light at focal lengths of 6.1 µm, 26.4 µm, and 60.1 µm, as the central 2, 4, or 6 elements are amorphized to implement a different phase pattern on the transmitted light. The focal point is defined here as the location with the maximum local intensity averaged within a 1.5µm × 1.5µm box, and the coordinate system is indicated in the inset of Fig. 4a. We note that in simulation the phase shift was adjusted to $1.75\pi$, as it matched the experimental result the best.

We then measured the transmitted field intensity for the same metasurface configurations at a wavelength of 1240 nm along the optical axis using a motorized stage[43] (see Method). Fig. 4b shows the measured intensity profiles, which clearly exhibit a change in the focusing behaviour as the metasurface is set to the respective phase profiles. Importantly, from measurements we obtain maximum field intensities along the optical axis at z=6.6 µm, 25.7 µm, and 62.5 µm, closely matching the simulation.

We note that, although the longitudinal intensity profiles closely match in experiment and simulation, the width of the focal spot in experiment appears more divergent. We primarily attribute this to the relatively broad linewidth of our laser source of ~1 nm (Supplementary Fig. S13), which smears out the optical phase contrast over a wider phase range. However, this issue can be directly addressed by employing a tunable laser source with narrower linewidth, which are commercially available (~ 1 pm).

To emphasize the change in the field intensity and the close match between simulation and experiment, Fig. 4c shows the intensity distributions along the optical axis (white dashed lines in Figs. 4a and 4b). We note that this axis is tilted in the experimental result because of the small incident angle (~3º) necessary to excite the guided-mode resonance (Supplementary Fig. S7). Besides the closely matching focal spots, we also observed abnormal intensity maxima at around 10µm in the 6-meta-molecule configuration (green line in d), which are attributed to strong near-field interactions.

While the tunable focusing demonstration is proof of the functionality for the device, further improvement can be directly achieved by increasing the pixel number of future iterations. Ultimately, this will extend the far-field beam shaping scheme to more complex tasks, such as beam steering and vortex beam generation, as a larger number of controllable meta-molecules becomes available.

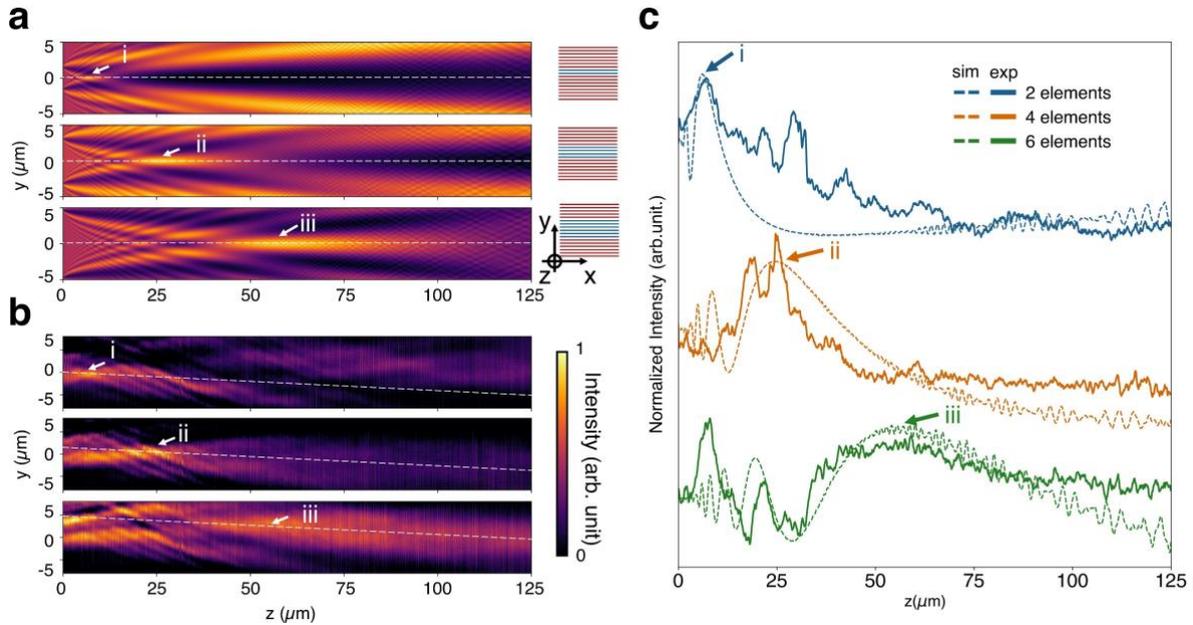

**Figure 4: Tunable far-field beam shaping.**

**a** Angular spectrum propagation simulation for a metasurface with 2 (top), 4 (middle), and 6 (bottom) central meta-molecules being amorphous, and the rest crystalline. The schematic on the right shows the corresponding phase mask (red: crystalline, blue: amorphous). The coordinate system is indicated below the schematics, with z being the optical axis. **b** Experimentally measured beam profiles. The plotted data is normalized to the globally crystalline configuration to remove the effect of objective focusing and then averaged in the *x* direction over the central 10 pixels of the metasurface. The focal points in **a** and **b** are denoted with white arrows and labelled from i to iii. The focal points are defined as the location with the highest local intensity averaged within a $1.5 \mu m \times 1.5 \mu m$ box. The near-field maxima below z=50µm are excluded while extracting the focal point in the last figure in the experiment. The white dashed lines denote the optical axis and experimental optical axis is tilted downward because of the 3° angle to excite the resonance mode. **c** Simulated (dashed lines) and measured (solid lines) longitudinal intensity distribution along the optical axis. The 2-, 4-, and 6-element configurations are shown by "blue", "orange" and "green" lines, respectively. All three configurations show a local intensity peak at their focal planes, marked with i to iii, corresponding to in **a** and **b**. The experimental result agrees well with the simulation. The plotted curves are smoothened using a 21-point moving average filter. (sim: simulation; exp: experiment; arb. unit: arbitrary unit)

**Conclusion**

In this work, we developed an electrically reconfigurable, transmissive metasurface that enables optical phase modulation in a non-volatile fashion and showcases individual pixel controllability, making it a spatial light modulator. We experimentally demonstrate a strong phase-only modulation (~0.2π) with less than 10% change in intensity, enabled by the coupling of a thin layer of low-loss phase-change material $Sb_2Se_3$ to a high-$Q$ metasurface. The achievable phase modulation is ten times larger than for a blanket layer without coupling to a metasurface. The device is robust against switching for over 1,000 times without degradation in performance. We further exploit a different resonance mode to show a much large resonance shift (~8nm), and 2π phase shift with individual meta-molecule control, supporting ten highly deterministic resonance levels. Finally, tunable far-field beam shaping with three different focal lengths is demonstrated by imparting different phase profiles, showcasing the capability of our SLM for free-space light manipulation.

Although we are limited to operation with distinct phase values at a specific wavelength, further work will explore full 2π phase-only control by carefully designing an avoided crossing between an over-coupled resonance and a resonance with high frequency tunability[38]. Huygen's metasurfaces with spectrally overlapped magnetic and electric dipole resonances can also be used to achieve 2π phase shift with unity transmission[26].

Overall, our work paves the way for non-volatile transmissive phase SLMs with zero static power consumption. The inherent memory of our device is crucial to simplify the electronic circuits for individual-pixel control. The thinness of the pixels (silicon and $Sb_2Se_3$, ~ 240 nm) further boasts a low mutual pixel crosstalk, hence a compact pixel size and a large field of view. Ultimately, we envision that with more independently controllable pixels, this device holds promise for various applications, such as semi-static display[44], structured light generation[45,46], free-space optical neural networks[47] and AR/VR[48].

**Methods**

*Device fabrication:* The reconfigurable diatomic metasurface is fabricated on a 220-nm thick silicon layer on top of a 3-μm-thick buried oxide layer (SOITECH) with back-side-polished silicon. The blanket SOI wafer is first implanted by phosphorus ions with a dosage of $2 \times 10^{15}$ cm$^{-2}$ and ion energy of 40 keV at a tilt angle of 7°. Subsequently, the wafer is annealed at 950 °C for 30 min to activate the dopants. The metasurface pattern is defined by a JEOLJBX-6300FS 100kV electron-beam lithography (EBL) system using positive tone ZEP-520A resist. 220 nm fully etched gratings are made by an inductively coupled plasma reactive ion etching (ICP-RIE) process in Florine-based gases. Before metallization, the surface native oxide was removed by immersing the chips in 10:1 buffered oxide etchant (BOE) for 15 s to ensure Ohmic contact. A second EBL exposure using positive tone poly(methyl methacrylate) (PMMA) resist is subsequently carried out to create windows for the Ti/Au deposition. After development, 5nm Ti followed by 150 nm Au was electron beam evaporated onto the chip. The lift-off of Ti/Au was completed again by immersing the chip in methylene chloride. Note that for the individual control of meta-molecules we replaced Au with Pt for electrodes to avoid the melting of the traces at high voltages. The third EBL step is used to expose the PMMA resist before depositing $Sb_2Se_3$ via magnetron sputtering. The $Sb_2Se_3$ is sputtered using a magnetron sputtering system at 30 W RF power under a deposition pressure of 4 mTorr and Ar flow of 30 sccm. The deposition rate for $Sb_2Se_3$ is ~1 nm/min. Additionally, the samples are capped with 10 nm of $SiO_2$ sputtered in situ (150 W RF power, 4 mTorr pressure, and Ar flow of 30 sccm), to prevent oxidation during sample shipping. The atomic ratio of $Sb_2Se_3$ after deposition is confirmed using XPS to be Sb:Se ≈ 44:56 which is close to the sputtering target stoichiometry of Sb:Se ≈40:60. Immediately after lifting off the PCM in methylene chloride, a 40nm ALD $Al_2O_3$ is grown on the chip to protect the PCM from oxidation and reflowing during switching. To allow good adhesion between the wedge bonds and the metal pads, the

5th EBL step is used to open windows in PMMA resist at the wire bonding regions for $Al_2O_3$ etching. The $Al_2O_3$ on top of the contacts is etched away using ICP-RIE etching in Chlorine-based gases. Then the PCMs are initialized into the fully crystalline state by rapid thermal annealing (RTA) at 200°C for 10 min under $N_2$ atmosphere before measurements. Finally, the chip is wire bonded onto the custom-made PCBs using a wire bonder (Westbond) via gold ball-wedge bonds.

*Optical spectral measurement:* The metasurface is characterized by a custom-build two-objective transmission setup (see Fig. 2b and Supplementary Fig. S14a). A collimated broadband laser (Fianium) is linearly polarized and focused onto the metasurface at normal incidence by a 10× near IR objective (Mitutoyo, 0.26 numerical aperture) and collected by a 50× near IR objective (Olympus, 0.65 numerical aperture). The beam splitter (Transmission: Reflection~85:15 in near IR) splits the light into two paths: one path (~85%) goes to the IR spectrometer (Princeton Instrument) for characterizing the optical spectrum; the other path (~15%) goes to the IR camera (NIT SenS 320) for imaging the metasurface. Supplementary Fig. 14b shows the modified setup for characterizing the cyclability of the tunable mesurface and beam splitting. A tunable continuous-wave laser (Santec TSL-510) is used here as the source to park the laser wavelength near the 1518nm resonance while the real-time transmission is monitored by a low-noise power meter (Keysight 81634B). To extract the *Q* factor, the diatomic resonance is fitted by a Fano line shape using the equation $T = | a + jb + \frac{c}{E-E_0+j\gamma} |^2$, where *a*, *b*, and *c* are constant real numbers. *E* is the photon energy and $E_0$ is the central resonance energy. 2γ is the line width of the resonance. So, the *Q* factor is calculated to be $\frac{E_0}{2\gamma}$.

*Optical phase measurement:* To measure the phase shift caused by the switching, we built a Mach-Zehnder interferometer (Supplementary Fig. S14b) and switched the pixel in-situ. The interference fringes between the signal beam through the metasurface and the reference beam are taken by the IR camera. The images are captured at 11 different wavelengths from 1513nm to 1523nm for the two optical states, averaged over 20 frames. The phase of each optical state is calculated by first applying a high pass filter on the image in the Fourier domain, and then take the argument of the filtered image in the real space. The phase shift is the difference between the extracted phase in crystalline state and amorphous state.

*Electrical control for individual addressing:* Each metasurface pixel on the chip is connected to a metal pad that is wire bonded to a pin on the carrier PCB or PCB1. A hole (0.8cm in diameter) is opened at the center of PCB1 to allow direct light transmission through the PCB. To individually control the pixels, a second customized PCB (PCB2) carrying a microcontroller (Arduino Nano) is inserted into the predefined pins on PCB1. Most commercial microcontrollers normally supply low voltage ($\leq 5V$) and slow speed (tens of megahertz), whereas high amplitude ($>10V$) and short (nanosecond falling edge) pulses are required to amorphized the PCMs. Hence the microcontroller cannot be directly used to switch PCMs, instead an external function generator is connected to the PCB2 as a source of excitation. The microcontroller can be programmed to turn on/off 17 MOSFETs connected in series with the on-chip microheaters. By switching the MOSFET, the voltage drop across the microheater can be controlled independently when an electrical pulse is applied. The resistance of the microheaters is measured using a source meter (Keithley 2450). A $30\mu m \times 30\mu m$ large metasurface pixel typically has resistance of ~115 Ω at 1mV DC bias, with ~ 45 Ω due to the relatively long Pt wires. The resistance of a single channel is ~1.45 kΩ. The SET and RESET pulses were generated from an arbitrary function generator (Keysight 81150A). To reconfigure the metasurface, we used a voltage pulse of 15 V (~10 mA/channel), 1.25 μs pulse width, and 8 ns rising/trailing edge to induce the amorphization. For crystallization, multiple voltage pulses of 6 V (~4.1 mA/channel), 50 μs pulse width, and 30 μs trailing edge is used. Our current device size is limited by the maximum voltage (20V at 50Ω load impedance) of our function generator. A larger aperture size will require function generator that can source higher voltage to amorphize the PCMs.

*Beam focusing simulation and measurement:* We used a 2D angular spectrum propagation program for the beam focusing simulation. A uniform input field distribution was assumed that imparts on the metasurface with centre 2, 4, 6 meta-molecules with $1.75\pi$ phase and 0 everywhere else. After transmission through the metasurface, the beam is numerically propagated in the momentum space and transformed back to real space to obtain the intensity distribution at a specific *z* position. In experiment (see Supplementary Fig. S14c), a broadband laser source (SuperK FIANIUM) with a tunable multi-channel filter (SuperK SELECT) is used to provide light near 1230 nm. To further reduce the linewidth, a dispersive grating was used before coupling the laser to the optical fibre which reduced the linewidth from ~15 nm to ~1 nm. The laser is collimated at the other end of the fibre, filtered by an iris, focused by a 10× NIR objective (Mitutoyo, 0.26 numerical aperture) on the metasurface to provide enough intensity. An imaging system, consisting of a 50× near IR objective (Olympus, 0.65 numerical aperture), a lens (Thorlabs AC254-200-C, $f$=200 mm), and an IR camera is on a $z$ motor stage. The raw data are videos (PTW files) taken by the IR camera with an exposure time of 1 ms and a frame rate of 50 frames/second while the motor stage moves away from the metasurface imaging plane with a velocity of 5 µm/second. All frames are then processed in MATLAB to obtain the cross-sectional intensity profile at each $z$-step. The results are normalized to the global crystalline metasurface data to compensate for the slight focusing due to the 10× objective.

**Authors contribution**

**Acknowledgements**

The research is funded by National Science Foundation (NSF-1640986, NSF-2003509), ONR-YIP Award, DRAPER Labs, DARPA-YFA Award, and Intel. F.M. is supported by a Draper Scholars Program. Part of this work was conducted at the Washington Nanofabrication Facility/ Molecular Analysis Facility, a National Nanotechnology Coordinated Infrastructure (NNCI) site at the University of Washington with partial support from the National Science Foundation via awards NNCI-1542101 and NNCI-2025489.

# Supplementary Information

# Non-volatile Phase-only Transmissive Spatial Light Modulators

Zhuoran Fang[1,+,*], Rui Chen[1,+,*], Johannes E. Fröch[1], Quentin A. A. Tanguy[1], Asir Intisar Khan[3], Xiangjin Wu[3], Virat Tara[1], Arnab Manna[2], David Sharp[2], Christopher Munley[2], Forrest Miller[1,4], Yang Zhao[1], Sarah J. Geiger[4], Karl F. Böhringer[1,5], Matthew Reynolds[1], Eric Pop[3], Arka Majumdar[1,2*]

[1]*Department of Electrical and Computer Engineering, University of Washington, Seattle, WA 98195, USA*

[2]*Department of Physics, University of Washington, Seattle, WA 98195, USA*

[3]*Department of Electrical Engineering, Stanford University, Stanford, CA 94305, USA*

[4]*The Charles Stark Draper Laboratory, Cambridge, MA 02139, USA*

[5]*University of Washington, Institute for Nano-engineered Systems, Seattle, Washington, United States*

[+]*These authors contributed equally*

*Email: arka@uw.edu, rogefzr@uw.edu, charey@uw.edu

S1. Electric field profile of the diatomic gratings in near field
S2. Extraction of the $Q$ factor
S3. GST phase change
S4. Temporal response of the reversible switching
S6. Broadband tunable notch filter based on GST
S7. Electrical wire design for equal resistance
S8. Transverse-electric polarized guided-mode resonance with a full $2\pi$ phase modulation
S9. Thermal crosstalk analysis
S10. Transmission spectra of four different programmed configurations
S11. The fitted Q factor with respect to the number of switched meta-molecules
S12. The spectrum of a narrowband laser source combining Fianium Supercontinuum laser and a grating filter
S13. Optical setup for transmission measurements

**S1. Electric field profile of the diatomic gratings in the near field**

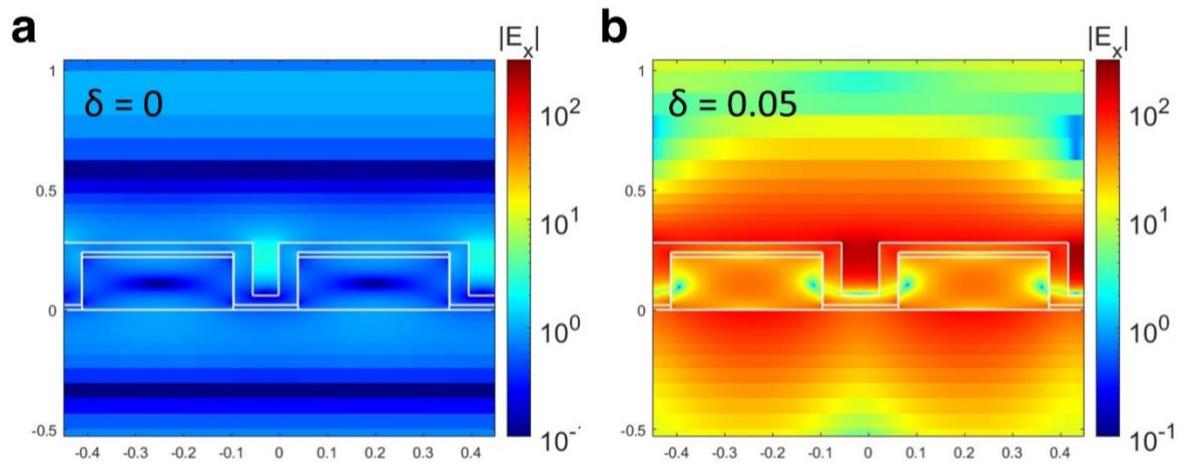

**Figure S1: Electrical field amplitude of the diatomic resonance.**

**a** The electric field profile when δ=0, corresponding to the high contrast gratings. **b** The field profile of diatomic metasurfaces on resonance. The unit for the scale bar is micrometer. Λ=900nm and Γ=0.7.

## S2. Extraction of the $Q$ factor

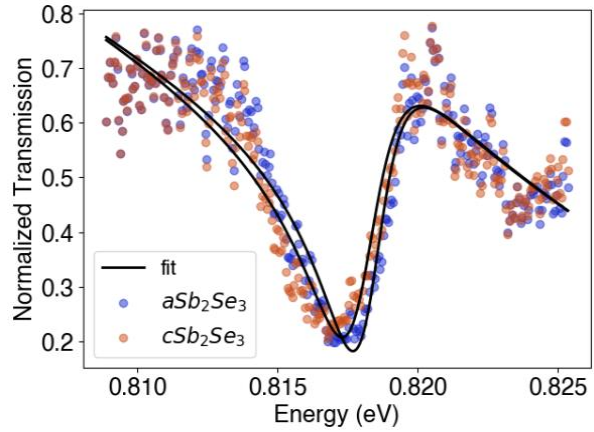

**Figure S2: Fitting the resonance in amorphous and crystalline state.**

The solid line shows the fitting result of the measured spectrum (dots) by a Fano line shape using the equation $T = |a + jb + \frac{c}{E - E_0 + j\gamma}|^2$, where $a$, $b$, and $c$ are constant real numbers. $E$ is the photon energy and $E_0$ is the central resonance energy. The spectrum is averaged over five switching cycles.

## S3. GST on the diatomic gratings

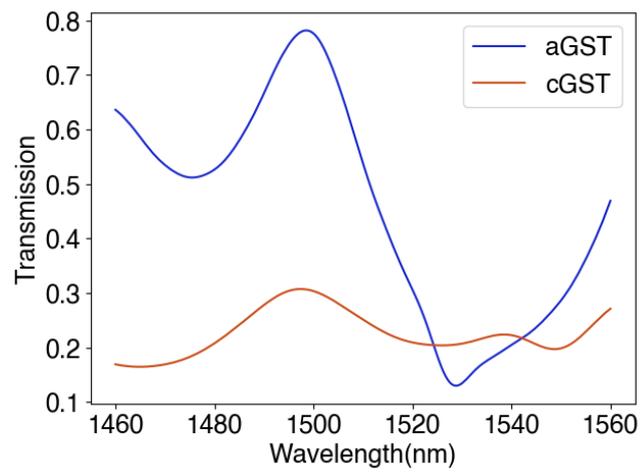

**Figure S3: Spectrum after capping 20nm GST on the diatomic gratings.**

The grating design is the same as in S1. Only the $Sb_2Se_3$ has been replaced by GST. This shows the effect of large absorptive loss from GST.

## S4. Temporal response of the reversible switching

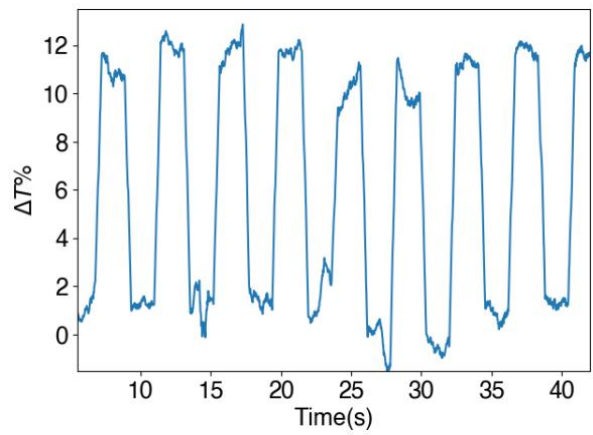

**Figure S4: Temporal response of the reversible switching.**

The switching conditions are 3.6V, 50μs pulse width, 30μs trailing edge for SET and 11.6V, 600ns pulse width, 8ns trailing edge for RESET.

## S5. Optical micrographs of the metasurface pixel before and after the cyclability test

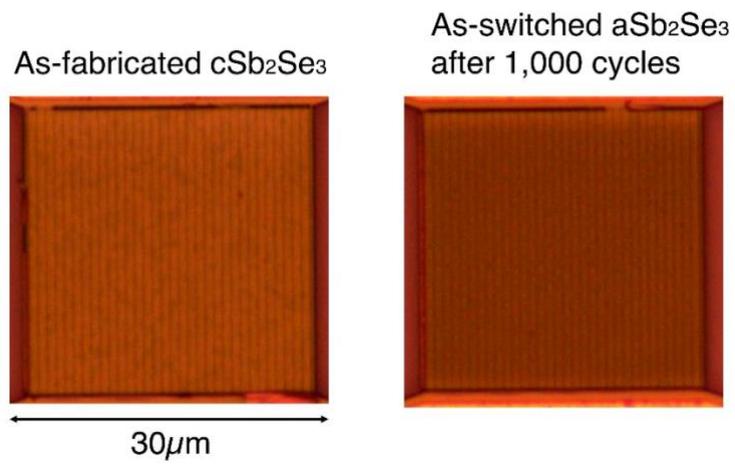

**Figure S5: Optical micrographs of the metasurface pixel before and after the cyclability test.**

The uniform color change across the entire metasurface indicates a uniform switching.

## S6. Reversible switching of blanket GST films using doped Si heater

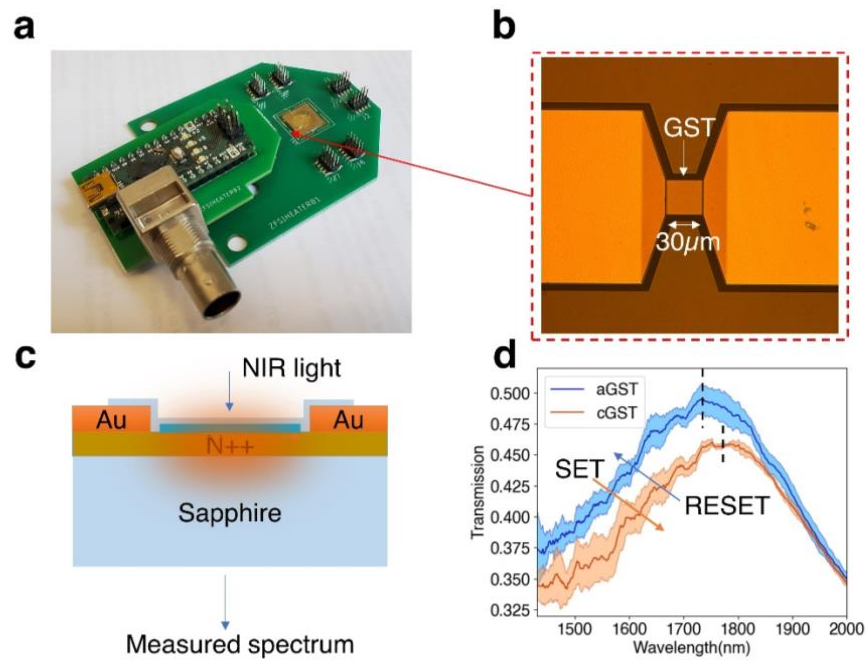

**Figure S6: Reversible switching of GST blanket film using doped Si heater.**

**a** A silicon-on-sapphire chip wire bonded to a customized PCB. **b** A switchable GST pixel on the chip. **c** Schematic of the device structure. The Si is 500nm thick and the GST is 20nm thick. **d** Reversible switching of the transmission spectrum of a 10μm large GST pixel measured by a FTIR. The spectrum is averaged across five cycles. 2.5V, 50μs wide, 50μs trailing edge pulses are used for crystallization, 7V, 200ns wide, 8ns trailing edge pulses are used for amorphization.

## S7. Electrical wire design for equal resistance

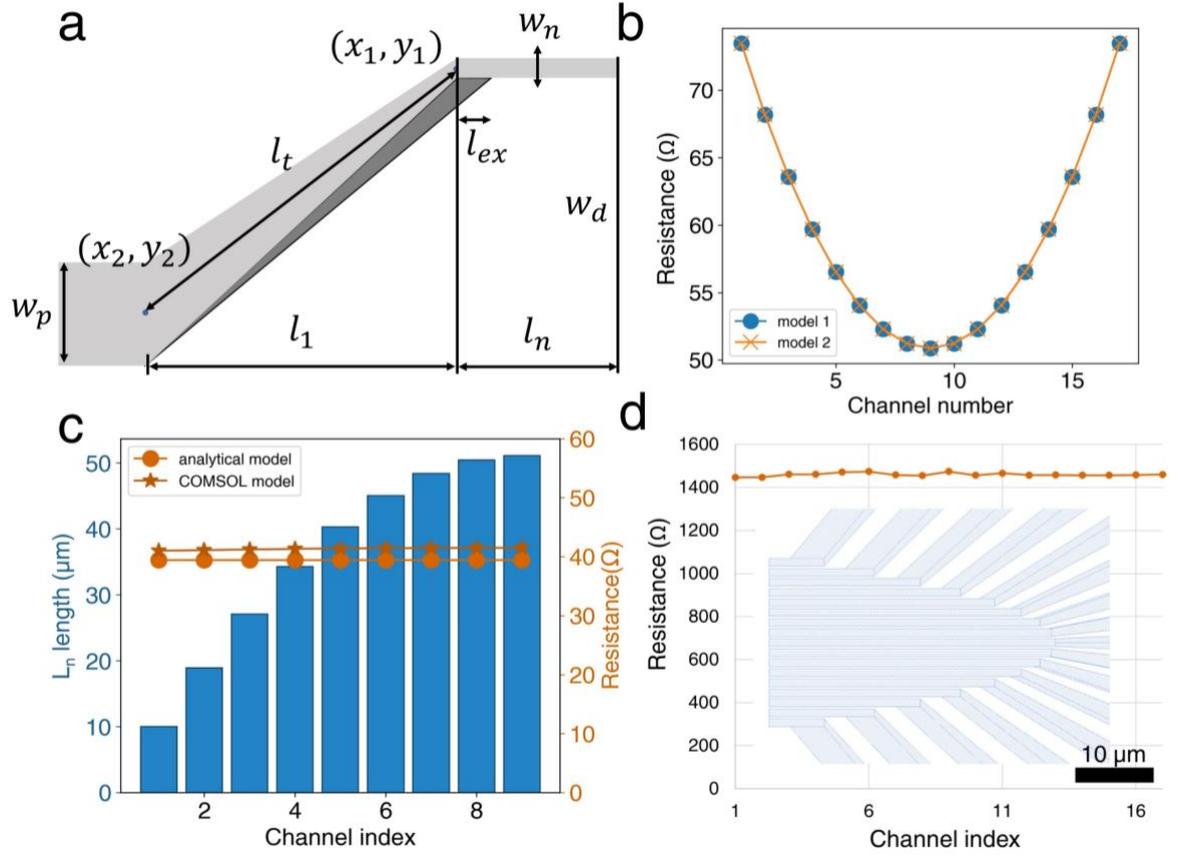

**Figure S7: Metal wire engineering to achieve equal resistance across 17 channels.**

**a** Schematic for a single metal wire with parameters denoted. 17 channels have different $w_d$, hence geometry engineering is necessary to achieve uniform resistance. For the first model, only the light gray region is used. In the second model, the dark gray region is added to avoid narrow width at the turn, enhancing the wire's robustness against high current. **b** Calculated resistance using model 1 (blue) and model 2 (orange), showing matching results of a large resistance variation of around 25Ω across the channels if $L_n$ is constant. **c** Optimized design for uniform resistances by varying $L_n$ across the channels (blue). The orange line with circle (star) markers is the resistance calculated from the analytical model (COMSOL simulation). **d** Experimentally measured resistance across 17 channels.

In this section, we derive an analytical model for the metal wires from the metasurface to the metal pads, depicted in Fig. S7a. The model was then used to design the metal wire geometry to obtain uniform resistance, which is essential for the simultaneous control of multiple channels. We denote the resistance of the rightmost straight wire as $R_1$ and that of the tilted part from points $(x_1, y_1)$ to $(x_2, y_2)$ as $R_2$. The pad resistance is usually small and is ignored. Therefore, the total resistance is calculated as $R = R_1 + R_2$.

In our first model, where the dark gray part is not considered, $R_1$ is given by:

$$R_1 = \frac{\rho l_n}{h w_n}$$

(1)

Where $\rho = 10.8 \times 10^{-8} \Omega/m$ is the resistivity of Pt, $h = 250 nm$ is the metal wire thickness, $l_n$ and $w_n$ are denoted in the schematic. The tilted wire can be approximated by a trapezoid, the resistance of which could be obtained by segmenting the wire into small rectangular pieces and integrating them over the entire length. Although this method ignores the boundary effects of electric currents, it works very well for a slowly varying shape. Considering an isosceles trapezoid with a top width of $w_t$, bottom width of $w_b$ and height of $l_t$, the resistance is calculated as $R = \frac{\rho}{ha_0} \ln\left(\frac{a_0 l_t}{w_t} + 1\right)$, where $a_0 = (w_b - w_t)/l_t$. Here, $w_t$ and $w_b$ could be approximated as:

$$\begin{cases} w_t = w_n \cdot \cos(\theta) \\ w_b = w_p \cdot \cos(\theta) \end{cases}$$

(2)

Where $w_n$ and $w_p$ are constants as denoted in the schematic, and $\theta$ is the tilting angle (0~π/2) of the wire, given by

$$\theta = \left|\arctan\left(\frac{y_2 - y_1}{x_2 - x_1}\right)\right|$$

(3)

Where $|a|$ means taking the absolute value of $a$. The length of the tilted wire can be obtained by $l_t = \sqrt{(x_1 - x_2)^2 + (y_1 - y_2)^2}$. Therefore, the total resistance is given by the following equations:

$$\begin{cases} R = \frac{\rho l_n}{h w_n} + \frac{\rho}{h a_0} \ln\left(\frac{a_0 l_t}{w_t} + 1\right) \\ a_0 = (w_b - w_t)/l_t \\ w_t = w_n \cdot \cos(\theta) \\ w_b = w_p \cdot \cos(\theta) \\ \theta = \left|\arctan\left(\frac{y_2 - y_1}{x_2 - x_1}\right)\right| \\ l_t = \sqrt{(x_1 - x_2)^2 + (y_1 - y_2)^2} \end{cases}$$

(4)

In the experiment, we observed that the corners were the most prone to damage. This is attributed to a narrow wire at the corner, especially for edge channels with large tilting angles.

We add extra metal wires drawn in dark gray to improve the robustness. The second model for this revised structure is given by:

$$\begin{cases} R = \dfrac{\rho(l_n - l_{ex}/2)}{hw_n} + \dfrac{\rho}{ha_0} \ln\left(\dfrac{a_0 l_t}{w_t} + 1\right) \\ a_0 = (w_b - w_t)/l_t \\ w_t = w_n \cdot \cos(\theta) + l_{ex} \cdot \sin(\theta) \\ w_b = w_p \cdot \cos(\theta) \\ \theta = \left|\arctan\left(\dfrac{y_2 - y_1}{x_2 - x_1}\right)\right| \\ l_t = \sqrt{(x_1 - x_2)^2 + (y_1 - y_2)^2} \end{cases}$$

(5)

As a sanity check, Eq. (5) reduces to Eq. (4) if $l_{ex} = 0$. The accuracy of our model is also verified by Fig. S7b, where two models match well. It can also be seen that there is a large resistance variation of ~25Ω across the 17 channels if we use the same $L_n$. By varying the $L_n$ spatially, it is possible to obtain uniform resistance (variation ~ ±0.5Ω, < 1%) across the channels as shown in Fig. S7c. The analytic model also agrees with the COMSOL simulation with less than 2Ω difference. The optimized metal wire shape is shown in the inset of Fig. S7d. In experiments, the doped-silicon gratings contribute to a much higher resistance than the simulation with an average resistance of ~1460Ω across 17 channels, see Fig. S7d. However, the variation in the resistance is only around ±10Ω (~1% of the resistance) which will not lead to significant variation in Joule heating.

## S8. Transverse-electric polarized guided-mode resonance with a full 2π phase modulation

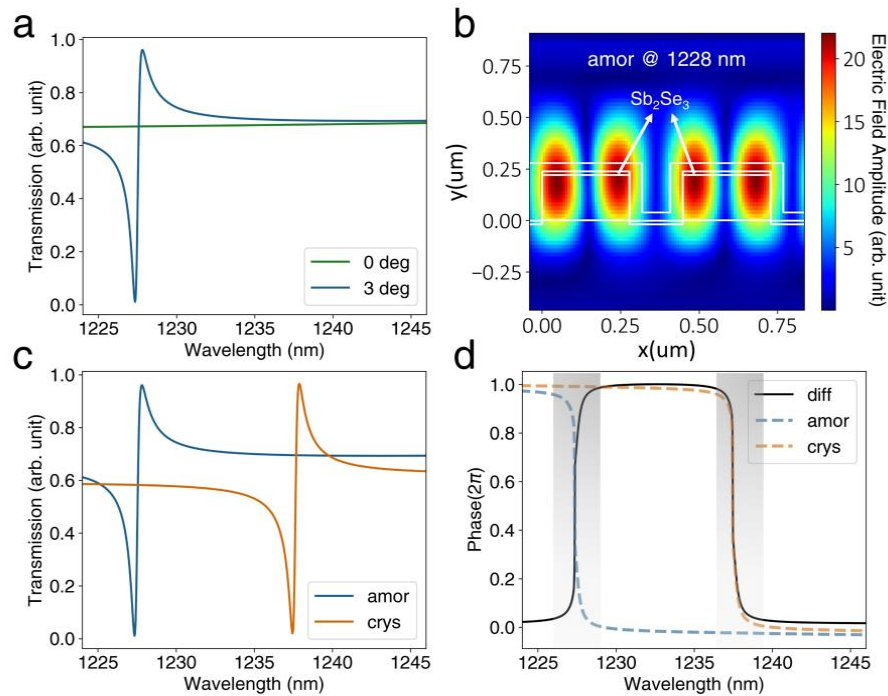

**Figure S8: Transverse-electric polarized guided-mode resonance mode with strong field overlap with $Sb_2Se_3$, achieving larger resonance shift and a full 2π phase modulation.**

**a** The transmission spectra for light with normal incident (green) and with a 3° angle (blue). The resonance is only found with a non-zero incident angle. We note that the resonance is not clearly visible below 3° and a further increase of the angle leads to a decrease in Q-factor. Therefore, we chose 3° because it supports a resonance mode with a relatively high Q-factor. The following results are simulated with an incident angle of 3°. **b** Electric field distribution for amorphous-$Sb_2Se_3$ at resonance wavelengths 1228nm. The white contour represents the metasurface structure. The field on $Sb_2Se_3$ is much stronger compared to the quasi-BIC mode in Fig. S1. Using crystalline-$Sb_2Se_3$ gives a similar field profile at 1238nm due to the relatively thin $Sb_2Se_3$. **c** Calculated transmission spectra at amorphous (denoted "amor", blue line) and crystalline (denoted "crys", orange line) state. A large spectral shift of ~10nm is observed. **d** Wavelength-dependent phase difference (denoted "diff", black line) between the amorphous (blue dashed line) and crystalline (orange dashed line) state with a maximum phase modulation of full 2π. One can achieve an arbitrary phase shift (0~2π) near the resonances (near 1228 or 1238 nm, denoted by the gray boxes) by precisely controlling the laser wavelength.

## S9. Thermal crosstalk analysis

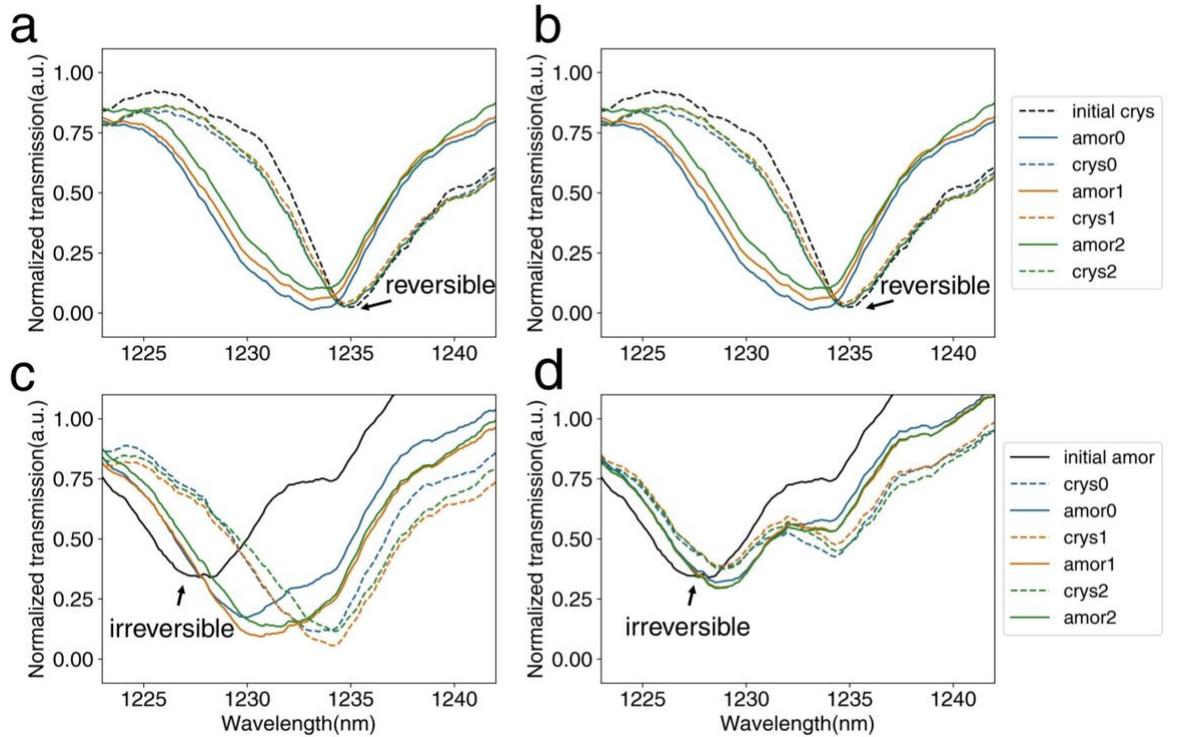

**Figure S9: Normalized transmission spectra showing evidence of thermal crosstalk when an individual meta-molecule is crystallized.**

Normalized transmission spectra for hybrid gratings with an amorphous period two and three times the original channel pitch (ii, iii in Fig. 3c). **a, b** We start from an all-crystalline state and reversibly switch the selected channels via amorphization. Different colors show multiple switching cycles. After applying the amorphization pulse, we can return the spectrum to the initial state every time using a crystallization pulse, as indicated by the overlapping between the dashed and black lines. **c, d** We start from an all-amorphous state and reversibly switch the selected channels via crystallization. The crystallization pulse leads to unintentional crystallization of the adjacent meta-atoms due to thermal crosstalk. As a result, the subsequent amorphization pulses cannot return the systems to the initial all-amorphous state and the switching is irreversible. In contrast, the amorphization pulses we used in **a, b** will not switch the adjacent channels due to a much smaller thermal mass and higher temperature threshold. We note a global amorphization pulse could bring the metasurface back to the initial all-amorphous state, ruling out the possibility of device damage.

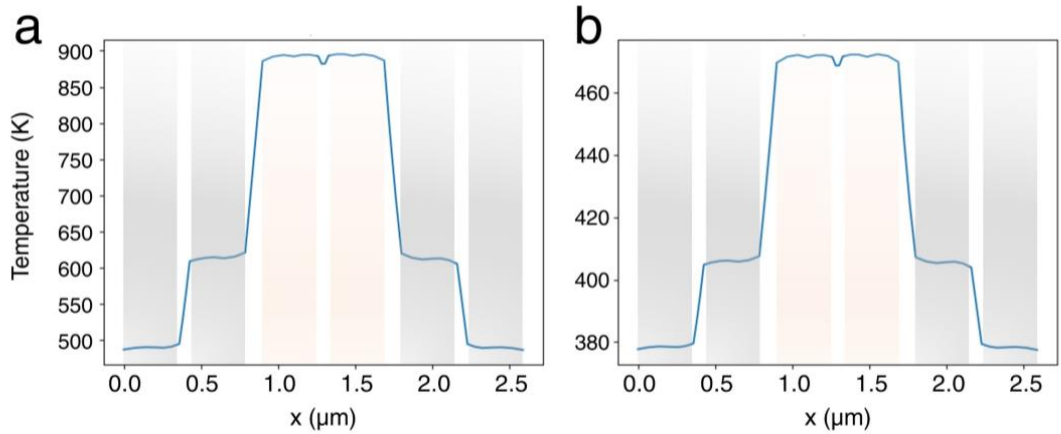

**Figure S10: Joule heating simulation in COMSOL.**

**a** Simulated temperature distribution for amorphization using a 10.8V, 1μs pulse. **b** Simulated temperature distribution for crystallization using a 4.5V, 50μs pulse. The grating structure is overlaid on the plot. The electric current is only injected in the center diatomic grating (light orange), while the adjacent gratings (gray) are heated up due to thermal crosstalk. The temperature drops quickly away from the center. The amorphization temperature drop by ~280 K on the nearest gratings due to the shorter pulse width. Since the adjacent gratings do not reach the melting point, the thermal crosstalk will not lead to inadvertent switching of the nearby cells. In contrast, the temperature only drops by ~60K for the nearest gratings after the crystallization pulse, which is not enough to suppress accidental crystallization.

**S10. Transmission spectra of four different programmed configurations**

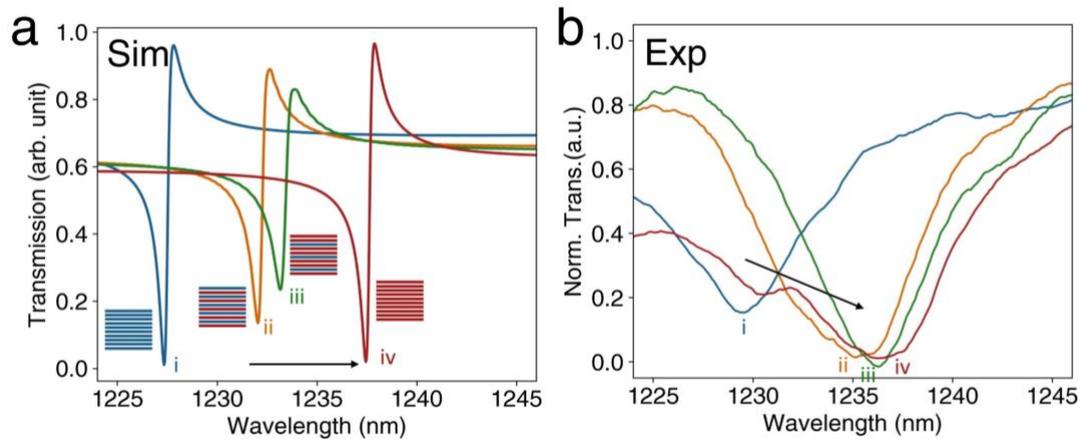

**Figure S11: Transmission spectra for four different programmed configurations in Fig. 3c.**

**a** Simulated spectra for four different grating configurations. A spectral red shift is observed from i to iv, corresponding to Fig. 3c: (i) metasurface with all amorphous meta-molecules; (ii),(iii) hybrid metasurface with an amorphous period two and three times the original channel pitch, respectively, and (iv) all crystalline metasurface. **b** Measured transmission spectral shift qualitatively agrees with the simulation. Sim: simulation; Exp: experiment.

**S11. The fitted Q factor with respect to the number of switched meta-molecules**

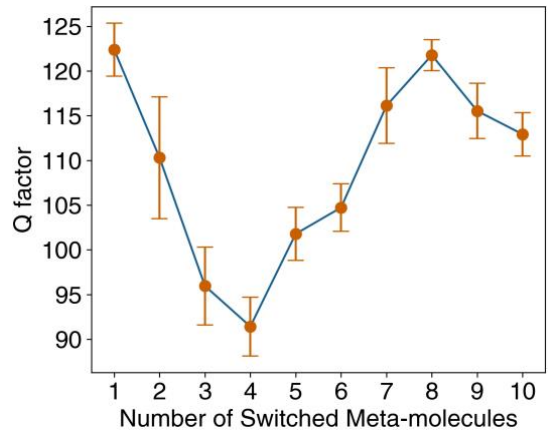

**Figure S12: Fitted Q-factor of the metasurface versus the number of switched meta-molecules.**

The Q-factor first decreases because of the inhomogeneous broadening of the amorphous-crystalline hybrid resonance mode and then increases as the amorphous resonance mode dominates. The Q reduction at the fully amorphous state near the end is not physical and could be attributed to a Fano resonance fitting error.

**S12. The spectrum of a narrowband laser source combining Fianium Supercontinuum laser and a grating filter**

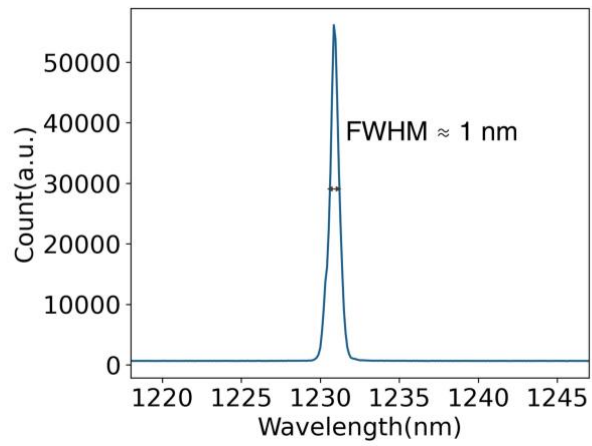

**Figure S13: Spectrum of the Fianium broadband source combined with a grating.**

## S13. Optical setup for transmission measurements

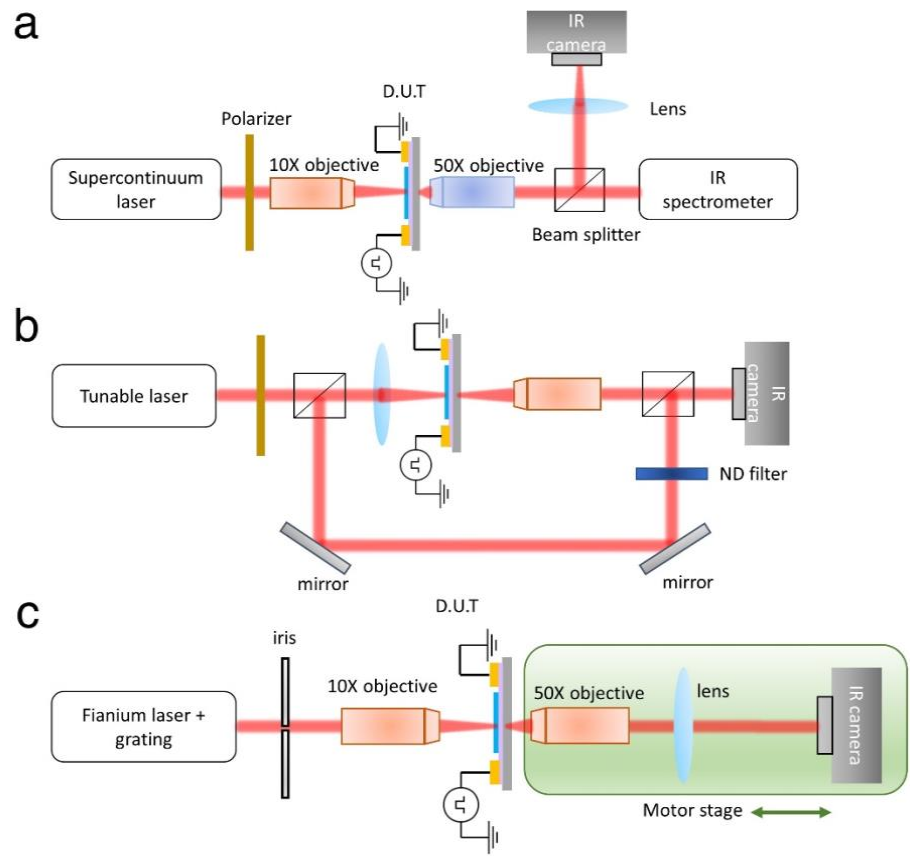

**Figure S14: Optical measurement setup with in situ electrical control of metasurface.**

**a** Transmissive spectral measurement setup. **b** Optical phase measurement setup based on a Mach Zehnder Interferometer. **c** Motorized stage setup for far-field beam profile imaging.